\documentclass{article}
\usepackage{graphics}
\title{$N$-component Bose-Einstein condensate in an
optical lattice:  destruction of the condensate and quasiparticle properties}

\author{Ch. Moseley and K. Ziegler \\
Institut f\"ur Physik, Universit\"at Augsburg, Germany \\
\small email: christopher.moseley@physik.uni-augsburg.de}

\begin{document}

\maketitle

\section*{Abstract}

We present a model of $N$-component hard-core bosons on a lattice. The limit
$N\rightarrow\infty$ can be solved exactly. A saddle point approximation leads
to a $1/N$ expansion and allows the calculation of physical quantities like the
density of the condensate, the correlation between the components and the
density-density correlation function and the correlation between different
components.
We find a superfluid phase with a tendency towards a Mott insulator at
high densities and finite $N$.

\section{Introduction}

A system of interacting lattice bosons undergoes a phase transition between
a superfluid and a Mott insulator \cite{fisher,ziegler93,jaksch}.
Motivated by the recent discovery of a Mott-insulating state in an
ultracold gas of Rubidium atoms in an optical lattice \cite{bloch},
we study the effect of strong (i.e. hard-core) interaction in a gas
of lattice bosons.
The strength of the interaction is tuned effectively by introducing $N$
components in the Bose gas [5-12].
These components can be considered as $N$ degenerate states per site
of the optical lattice, assuming that
there is a hard-core interaction only in the same state. Thus, the
bosons can avoid hard-core interaction by choosing an unoccupied state,
leading to an effective interaction between the bosons that becomes
weaker with an increasing number of states $N$.
Another tunable parameter is the tunneling rate $t$ between
neighboring sites of the optical lattice. For sufficiently small
tunneling rate the Bose-Einstein condensate disappear, since
for $t=0$ (atomic limit) there can be no condensate but only a (Mott)
insulator. In order to study the properties of such a multicomponent
Bose gas we evaluate physical quantities like the particle density,
the condensate density, the structural order parameter with respect to
the $N$ components and the density-density correlation function.
Our method is a saddle-point integration for large values of $N$ and
a $1/N$ expansion \cite{ziegler94}.

\section{The Model}

The Hamiltonian for a Bose gas in a $d$-dimensional
optical lattice with $\cal N$ sites and $N$ states per site is
\[
H=-\sum_{r,r'}\sum_{\alpha,\alpha'=1}^Nt_{r,\alpha,r'\alpha'}
a_{r,\alpha}^\dagger a_{r',\alpha'},
\]
where $a_{r,\alpha}^\dagger$ ($a_{r,\alpha}$) is the creation (annihilation)
operator of a hard-core boson at site $r$ on the optical lattice and in state
$\alpha$, i.e. $(a_{r\alpha}^\dagger)^2=0$. The states $\alpha=1,2,..., N$
refer to $N$ different components of the Bose gas. $t_{r,\alpha, r'\alpha'}$
is the tunneling rate for a
hard-core boson from the state $\alpha'$ at site $r'$ to the
state $\alpha$ at site $r$. Here we will assume isotropic tunneling
\[
t_{r,\alpha,r'\alpha'}=\cases{
J/N & for $r,r'$ nearest neighbors \cr
0 & otherwise \cr
}.
\]
For the functional integral we introduce
\[
(w_{r,r'})=e^{{\hat t}/2}
\]
with ${\hat t}=(t_{r,\alpha,r'\alpha'})$ (cf. \cite{ziegler02}).

This system can be characterized by a diagonal order parameter for
the components $\alpha$:
\[
a_{r,\alpha}^\dagger a_{r,\alpha}=n_{r,\alpha}
\]
which is the number operator of component $\alpha$. An off-diagonal
order parameter is
\begin{equation}
 a_{r,\alpha}^\dagger a_{r,\alpha'}\ \ \ (\alpha'\ne \alpha)
 \label{structop}
\end{equation}
which
measures how likely it is to switch a particle from component $\alpha'$
to component $\alpha$. Besides the expectation value of these
order parameters we also need their correlations to describe
ordering of these degrees of freedom.

\subsection{Functional Integral and Saddle-Point Approximation}

The starting point is the grand-canonical ensemble
of the $N$-state hard-core Bose gas. It is defined by the partition
function $Z_{HC}$
\[
Z_{HC}=\mbox{Tr} \, e^{-\beta (H -\mu \sum_r n_r)},
\]
where $-\mu \sum_r n_r$ controls the particle number in the grand-canonical
ensemble \cite{negele}.In the following, we will use the fugacity $\zeta=e^\mu$
instead of the chemical potential $\mu$.
It was shown \cite{ziegler94} that $Z_{HC}$ can be expressed
by a functional integral with respect to
two complex fields $\varphi_x$ and $\chi_x$. $x=(r,t)$ are coordinates of a
space-time lattice, where the time is $t=1,2,...,\beta$. The field is
periodic with respect to time: $\varphi_{\beta+1,r}=\varphi_{1,r}$.
The partition function then reads
\begin{equation}
 Z_{HC}=z_N
 \int \prod_x \frac{
 d \varphi_x d \varphi_x^\ast d \chi_x d \chi_x^\ast}{(2 \pi i)^2}
 e^{- S}
 \label{SP1}
\end{equation}
with the action
\[
 S :=
 \frac N\zeta \sum_{x,x'}({\bf 1}+\zeta w)^{-1}_{x,x'}\varphi_x
\varphi^\ast_{x'}
 +\frac N\zeta \sum_{x}|\chi_x|^2
\]
\begin{equation}
 - \sum_x\sum_\alpha \log [ 1 + (\varphi_x^\ast +i \chi_x^\ast )
(\varphi_x +i \chi_x ) ]
 \label{SP2}
\end{equation}
and the prefactor
\[
z_N=(\det \hat{v}^{-1}) N^{2\beta \cal N}.
\]
The number of states $N$ only appears in front of the
$N$-independent action in the partition function. Therefore
the functional integration can be performed for large values
of $N$ by a saddle-point (SP) approximation \cite{ziegler02}.
The corresponding SP equations can be solved with uniform solutions:
\begin{equation}
 \varphi_x \equiv \varphi, \ \  \chi_x \equiv \chi .
 \label{SP4}
\end{equation}
With $\sum_{x'} \hat{v}^{-1}_{xx'} = \frac{1}{2}$ two uniform
SPs are found, a trivial SP
\begin{equation}
  \varphi = \varphi^\ast = \chi = \chi^\ast = 0
 \label{SP5}
\end{equation}
labeled by $\nu=0$ and a non-trivial SP
\begin{displaymath}
 |\varphi| = 2 \sqrt{ 1-\frac{1}{\zeta} }
\end{displaymath}
\begin{equation}
 i\chi = -{1\over 2}\varphi,\ \ \ i\chi^\ast = -{1\over 2}\varphi^\ast
 \label{SP6}
\end{equation}
with broken $U(1)$ symmetry, labeled by $\nu=1$.
The complex fields $\varphi_x$ and $\chi_x$
are expected to fluctuate about the SP solution
\begin{equation}
 \varphi_x = \varphi + \delta\varphi_x \: , \:
 \varphi_x^\ast = \varphi^\ast + \delta\varphi_x^\ast \: , \:
 \chi_x = \chi + \delta\phi_x \: , \:
 \chi_x^\ast = \chi^\ast + \delta\chi_x^\ast \: .
 \label{SP7}
\end{equation}
In order to study the effect of the fluctuations, we
introduce a four component real field
\begin{equation}
 \delta\phi_x :=
 \left( \begin{array}{c}
 \delta\phi_{1,x} \\ \delta\phi_{2,x} \\
 \delta\phi_{3,x} \\ \delta\phi_{4,x}
 \end{array} \right) \equiv
 \left( \begin{array}{c}
 \delta\varphi'_x \\ \delta\varphi''_x \\
 \delta\chi'_x \\ \delta\chi''_x
 \end{array} \right).
 \label{SP8}
\end{equation}
Substituting this result into the partition function
leads to a Gaussian approximation
\begin{equation}
 Z_{HC,\nu} \approx Z_{\nu} Z_{QP,\nu},
 \label{SP15}
\end{equation}
where the first factor is the partition function for
$N \sim \infty$
\begin{equation}
 Z_0 =  z_N
\end{equation}
\begin{equation}
 Z_1 = z_N
 \exp \left[ -N\beta{\cal N} \left( 1- \frac{1}{\zeta}
- \log \zeta \right) \right]
 \label{SP16}
\end{equation}
and the second factor describes the partition function of
quasiparticles, which arise from $1/N$-corrections
\begin{equation}
 Z_{QP,\nu} = \int \prod_x \frac{d \delta \phi_x}{\pi^2}
 \exp \left[ -N \sum_{xx'} \langle \delta \phi_x, ({\cal G}_\nu^{-1})_{xx'}
 \delta \phi_{x'} \rangle \right],
 \label{SP17}
\end{equation}
where $\langle\ldots,\ldots\rangle$ denotes a scalar product.

The prefactor $\beta \cal N$ in Eq. (\ref{SP16}) comes from the summation over all
lattice sites $x$.
The Fourier components of the quasiparticle Green's functions are
$4 \times 4$-matrices of the form
\begin{equation}
 ({\cal G}_\nu^{-1})_{k\omega} =
  \left( \begin{array}{cc}
   \tilde{v}^{-1}_{k\omega} - A_\nu & -iA_\nu \\
   -iA_\nu                          & 1 + A_\nu
  \end{array} \right)
 \label{QP2}
\end{equation}
with the $2 \times 2$-matrices

\begin{displaymath}
 \tilde{v}^{-1}_{k\omega} =
 \frac{1}{B_k^2 + 1 + 2B_k \cos (\omega \tau)} \left(
  \begin{array}{cc}
   B_k \cos (\omega \tau) + 1 & B_k \sin (\omega \tau) \\
   -B_k \sin (\omega \tau)    & B_k \cos (\omega \tau) + 1
  \end{array} \right) ,
\end{displaymath}
$B_k=1-\tau J+\tau {J\over d}\sum_{j=1}^d\cos k_j$
and
\begin{displaymath}
 A_0 = \zeta
  \left( \begin{array}{cc}
   1 & 0 \\
   0 & 1
  \end{array} \right) \quad , \quad
 A_1 = \left( \begin{array}{cc}
   2/ \zeta -1 & 0 \\
   0           & 1
  \end{array} \right) \; .
\end{displaymath}
Now we can perform the Gaussian integral over the fluctuating field
(\ref{SP17}) and get
\begin{equation}
 Z_{QP, \nu} = N^{-2\beta \cal N} \prod_{k,\omega}
 \det ({\cal G}_\nu^{1/2})_{k\omega} \; .
 \label{QP3}
\end{equation}

\subsection{Spectrum of quasiparticles}
In the superfluid phase, the matrix $(\hat{\cal G}_1)_{k,\omega}$ can be recognized as
Green's function of quasiparticles, the elementary excitations out of the condensate.
The quasiparticle poles of the eigenvalues of $\det(\hat{\cal G}_1)_{k,\omega}$ lead to
the excitation spectrum. Identifying $i\omega$ with the excitation energies $\epsilon$,
one gets for small wave vectors $k$:
\begin{displaymath}
 \epsilon_k = \sqrt{\frac Jd (\zeta-1) k^2 +
 \left( \frac{J}{2d} k^2 \right)^2 }
\end{displaymath}
In the region $\zeta \approx 1$, where $\mu_0\approx 1+\zeta$ is
valid for the chemical
potential, one can identify $m\equiv d/J$ and gets the Bogoliubov spectrum for the weakly
interacting Bose gas
\begin{displaymath}
 \epsilon_k = \sqrt{\frac{\mu_0 k^2}{m} + \left( \frac{k^2}{2m} \right)^2} \; .
\end{displaymath}
The linear dispersion relation for small wave vectors $k$ is due to a Goldstone mode.

\section{Physical quantities}
\subsection{Correlation functions}
Correlation functions can be defined by introducing the generating fields
$\eta_x^{\alpha 1}$ and $\eta_x^{\alpha 2}$ and a generating functional
\begin{displaymath}
 Z_{gen}(\{ \eta_x^{\alpha 1} \},\{ \eta_x^{\alpha 2} \}) =
 z_N \int \prod_x \frac{
 d \varphi_x d \varphi_x^\ast d \chi_x d \chi_x^\ast}{(2 \pi i)^2}
 e^{- S_{gen}(\{\eta_x^{\alpha 1}\} , \{ \eta_x^{\alpha 2} \})}
\end{displaymath}
where
\begin{displaymath}
 S_{gen}(\{\eta_x^{\alpha 1} \} , \{ \eta_x^{\alpha 2} \} :=
 \frac N\zeta \sum_{x,x'}({\bf 1}+\zeta w)^{-1}_{x,x'}\varphi_x \varphi^\ast_{x'}
 + \frac N\zeta \sum_{x}|\chi_x|^2
\end{displaymath}
\begin{displaymath}
 - \sum_x\sum_\alpha \log [ 1 + (\varphi_x^\ast +i \chi_x^\ast + \eta_x^{\alpha 1})
 (\varphi_x +i \chi_x + \eta_x^{\alpha 2}) ] \; .
\end{displaymath}
$\eta_x^{\alpha 1}$ generates the field of the bosons and $\eta_x^{\alpha 2}$
the field of empty sites. According to Eqs. (\ref{SP1}) and (\ref{SP2})
the partition function is given as
\begin{displaymath}
 Z_{HC} = \left. Z_{gen}(\{ \eta_x^{\alpha 1} \},\{ \eta_x^{\alpha 2} \})
 \right|_{ \eta_x^{\alpha 1}\equiv\eta_x^{\alpha 2}\equiv 0} \; .
\end{displaymath}
An $n$-particle correlation function can be defined as
\begin{displaymath}
 C(x_1\alpha_1,\ldots,x_n\alpha_n ; {x'}_n\beta_n,\ldots,{x'}_n\beta_1) :=
 \left. \frac 1{Z_{HC}} \, \frac{\partial^{2n}Z_{gen}}{
 \partial\eta_{x_1}^{\alpha_1 1} \ldots \partial\eta_{x_n}^{\alpha_n 1}
 \partial\eta_{{x'}_n}^{\beta_n 2} \ldots \partial\eta_{{x'}_1}^{\beta_1 2} }
 \right|_{ \eta_x^{\alpha 1}\equiv\eta_x^{\alpha 2}\equiv 0} \; .
\end{displaymath}

All physical quantities that we present in the following are defined in
terms of these expectation values. The only exception is the total
density of particles, that can be defined as
\begin{equation}
 n:=1 - C(x\alpha;x\alpha).
 \label{ntotcorr}
\end{equation}
All physical results are evaluated for zero temperature.

\subsection{Total density and density of the condensate}
The total particle density per site and component (\ref{ntotcorr}) gives in
the large-$N$ limit
\begin{equation}
 n_\infty= \left\{
 \begin{array}{l@{\quad}l}
  0            & \mbox{if} \quad \zeta <1 \\
  1-\zeta^{-1} & \mbox{if} \quad \zeta >1
 \end{array} \right. \; .
 \label{ninfty}
\end{equation}
The $1/N$-correction of $n$ in the empty phase $\zeta<1$ vanishes but in the
superfluid phase the total density is enhanced. For large values of $\zeta$
the density approaches the value 1 (the ``full'' lattice), indicating the
tendency to a Mott insulating state.

Our definition of the condensate density is based on off-diagonal
long-range order, using the single particle correlation function \cite{leggett}:
\begin{equation}
 n_0 := \lim_{r-r'\rightarrow\infty} \frac 1\beta  \sum_{t=1}^\beta
 C((r,0)\alpha;(r',t)\alpha)
 \label{n0}
\end{equation}
For this quantity, the large-$N$ limit is
\begin{equation}
 n_0 = \left\{
 \begin{array}{l@{\quad}l}
  0                       & \mbox{if} \quad \zeta <1 \\
  \zeta^{-1} - \zeta^{-2} & \mbox{if} \quad \zeta >1
 \end{array} \right.
 \label{n0infty}
\end{equation}
and indicates a depletion for large values of $\zeta$.
The $1/N$-corrections even show
a destruction of the condensate density for sufficiently large $\zeta$, which
can be regarded as another hint at the formation  of a Mott insulator, where the phase
coherence and therefore the condensate is expected to be destroyed
(see figures 1 and 2).

Because of the Goldstone mode in the excitation spectrum of quasiparticles,
the $1/N$-corrections diverge for the dimensions $d=1$ and $d=2$, in agreement with the
expectation that in these cases there should not be a condensate.

\subsection{Structural order parameter}
The appropriate quantity describing the correlation of the internal structure of the
components of a given site is given by the expectation value of (\ref{structop}) as
\begin{equation}
 S_{\alpha,\beta} := C(x\alpha;x\beta) \; .
 \label{SOP}
\end{equation}
For $\alpha=\beta$ this is equal to $1-n$, $n$ being the total density of particles.
For $\alpha\neq\beta$ the $N\rightarrow\infty$-limit is equal to that of the condensate
density, but the $1/N$-correction is slightly higher. As for the condensate density,
the corrections indicate a destruction of the condensate
for large values of $\zeta$. Therefore this correlation function seems to be suppressed
in a Mott insulating phase, too (see figure 3).

\subsection{Density-density correlation function}
The relation (\ref{ntotcorr}) suggests the definition of a density-density
correlation function as
\begin{equation}
  {\cal D}^{\alpha\beta}(x,x') := 1 -  C(x\alpha ; x\alpha) -
  C(x'\beta ; x'\beta) + C(x\alpha,x'\beta ; x'\beta,x\alpha) \; .
\end{equation}
In the large-$N$ limit this correlation function reduces to the
square of the particle density:
\begin{displaymath}
 \lim_{N\rightarrow\infty} {\cal D}^{\alpha\beta}(x;x') = n^2
\end{displaymath}
The correlation length of spacially decaying $1/N$-corrections diverges at the
critical fugacity $\zeta=1$:
\begin{displaymath}
 \xi_{\cal D} \propto (\zeta-1)^{-\frac 12} \; .
\end{displaymath}

\section{Conclusions}

We have found a depletion of the condensate caused by strong interaction
in a dense $N$-component Bose gas. This effect leads eventually to the
destruction of the condensate and to the formation of a Mott insulator
for sufficiently large fugacity $\zeta$ at $N<\infty$. However, the condensate
survives in the limit $N\to\infty$, since the interaction is always weak in
this case. On the other hand, in the very dilute regime (i.e. for
$\zeta\approx 1$) the condensate density increases with decreasing $N$.
This reflects the well-known fact that an increasing interaction supports
BE condensation in a dilute Bose gas \cite{stoof92,ziegler00}.
The structural order parameter $S_{\alpha,\beta}$ measures the correlation
between the $N$ different states of the Bose gas. This quantity behaves
qualitatively like the condensate density but differently
on a quantitative level: the dilute regime, where the interaction supports this
parameter, is broader and its destruction happens at higher values of
$\zeta$.

Our saddle-point approximation is similar to the Bogoliubov theory at
low densities \cite{leggett} but deviates substantially from the latter at higher
densities. In particular, the appearence of a gap in the excitation spectrum of the
Mott-insulating phase is beyond the Bogoliubov theory.

\vskip0.5cm
\noindent
Acknowledgement: This work was supported in part by the Deutsche
Forschungsgemeinschaft through Sonderforschungsbereich 484.


\section*{Appendix}
\subsection*{Evaluation of correlation functions}
Correlation functions can be written as expecation values with respect to the
fields $\varphi,\chi$. The results for the functions mentioned in this work are:
\begin{eqnarray*}
 && C(x\alpha ; x\alpha) =
 \left\langle \left[1+\zeta(\varphi_x+i\chi_x)(\varphi_x^\ast+i\chi_x^\ast)\right]^{-1} \right\rangle \\
 && C(x\alpha ; x'\beta) =
 \left\langle \frac{\zeta(\varphi_x+i\chi_x)(\varphi_{x'}^\ast+i\chi_{x'}^\ast)}{
 [1+\zeta(\varphi_x+i\chi_x)(\varphi_x^\ast+i\chi_x^\ast)]
 [1+\zeta(\varphi_{x'}+i\chi_{x'})(\varphi_{x'}^\ast+i\chi_{x'}^\ast)]} \right\rangle
 \quad , (x,\alpha)\neq (x',\beta) \\
 && C(x\alpha,x'\beta ; x'\beta,x\alpha) =
 \left\langle \left[1+\zeta(\varphi_x+i\chi_x)(\varphi_x^\ast+i\chi_x^\ast)\right]^{-1}
 \left[1+\zeta(\varphi_{x'}+i\chi_{x'})(\varphi_{x'}^\ast+i\chi_{x'}^\ast)\right]^{-1} \right\rangle
 \quad , (x,\alpha)\neq (x',\beta)
\end{eqnarray*}
The functions can be expanded up to second order in the fields $\varphi$ and $\chi$, and
the SP approximation is applied. The zeroth order provides the large-$N$ limit
and the second order leads to a $1/N$-correction that appears as linear
combination of
two different expectation values:

\begin{displaymath}
 \Phi(r-r',t-t'):= \left\langle(\delta\varphi_x+i\delta\chi_x)
 (\delta\varphi_{x'}+i\delta\chi_{x'})\right\rangle =
 \left\langle(\delta\varphi_x^\ast+i\delta\chi_x^\ast)
 (\delta\varphi_{x'}^\ast+i\delta\chi_{x'}^\ast)\right\rangle \approx
\end{displaymath}
\begin{displaymath}
 \hspace{4cm}
 \frac 1N \int \frac{d^dk}{(2\pi)^d} \int_0^{2\pi} \frac{d\omega}{2\pi}
 \frac{B(k)^2 (\zeta-1)}{2B(k)\cos(\omega)-\zeta+B(k)^2 (\zeta-2)}
 \, e^{i[k(r-r')-\omega(t-t')]}
\end{displaymath}
\begin{displaymath}
 \hat\Phi(r-r',t-t'):= \left\langle(\delta\varphi_x+i\delta\chi_x)
 (\delta\varphi_{x'}^\ast+i\delta\chi_{x'}^\ast)\right\rangle \approx
 \hspace{5cm}
\end{displaymath}
\begin{displaymath}
 \hspace{4cm}
  \frac 1N \int \frac{d^dk}{(2\pi)^d} \int \frac{d\omega}{2\pi} \,
  \frac{B(k)(B(k)-\zeta\cos(\omega))}{2B(k)\cos(\omega)-\zeta+B(k)^2(\zeta-2)}
  e^{i[k(r-r')-\omega(t-t')]}
\end{displaymath}
with the function $B(k):=1-J+\frac Jd \sum_{j=1}^d \cos k_j$.

Spacially constant corrections depend on $\hat\Phi(0,0)$ and $\Phi(0,0)$ only.
The integral with respect to $\omega$ can be performed exactly and the $k$-integration
was done numerically.
\pagebreak
\begin{figure}
 \includegraphics{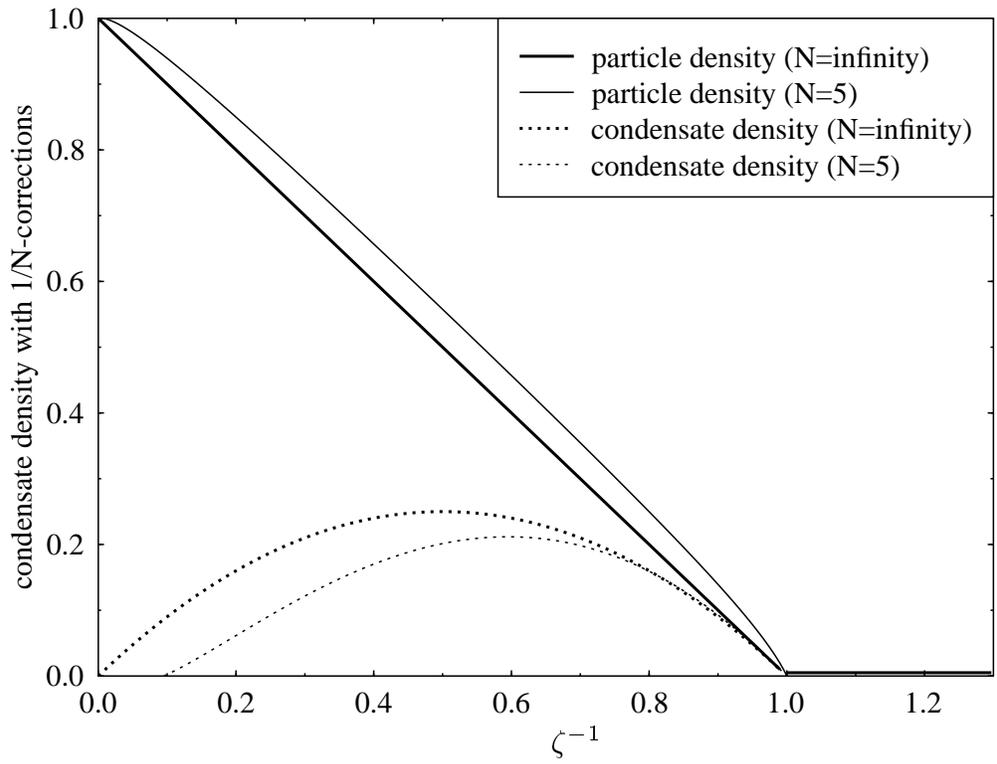}
 \caption{Total particle density and condensate density for $J=0.1$}
 \label{fign}
\end{figure}
\pagebreak
\begin{figure}
 \includegraphics{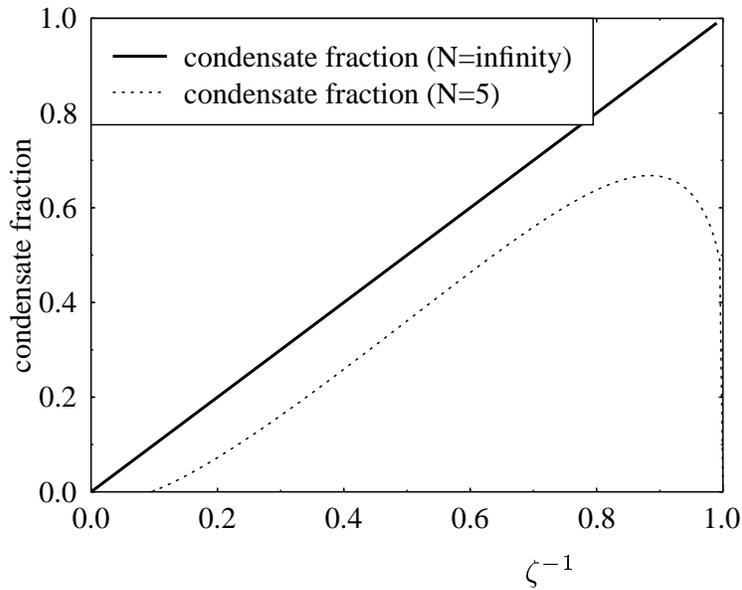}
 \caption{Condensate fraction $n_0/n$ for $J=0.1$}
 \label{figcf}
\end{figure}
\pagebreak
\begin{figure}
 \includegraphics{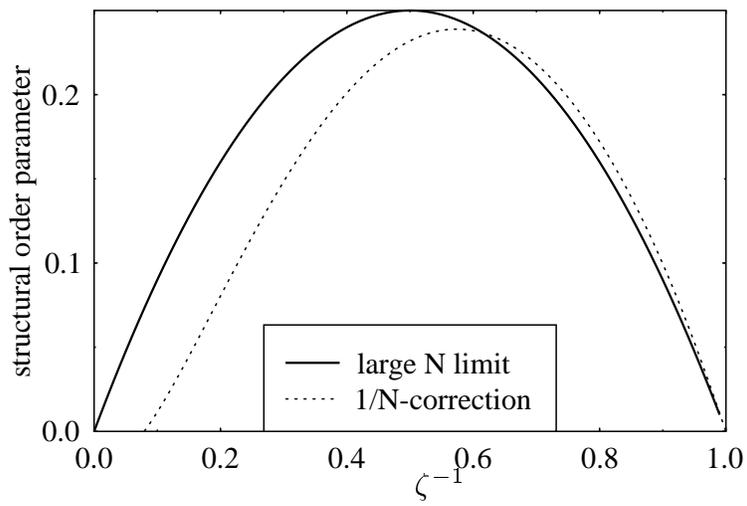}
 \caption{Structural order parameter for $J=0.1$}
 \label{figss}
\end{figure}

\end{document}